\documentclass[multphys,vecphys]{svmult}

\usepackage{makeidx}     % allows index generation
\usepackage{graphicx}    % standard LaTeX graphics tool
\usepackage{multicol}    % used for the two-column index
\makeindex             % used for the subject index
\begin{document}

\title*{Low frequency observations of SN 1993J with\\ 
Giant Meterwave Radio Telescope}
\titlerunning{SN 1993J low frequency spectrum with GMRT}

\author{P. Chandra\inst{1}\and
A. Ray\inst{2}
\and S. Bhatnagar\inst{3}}

\institute{Joint Astronomy Programme-Indian Institute of Science, Bangalore\\
 Tata Institute of Fundamental Research, Mumbai,
\texttt{poonam@tifr.res.in}
\and Tata Institute of Fundamental Research, Mumbai, \texttt{akr@tifr.res.in}
\and National Radio Astronomy Observatory, New Mexico, \texttt{sbhatnag@aoc.nrao.edu}}

\maketitle

\begin{abstract}

 In this paper, we discuss the low frequency spectrum of SN 1993J
with GMRT. We observed SN 1993J at several epochs
in 20cm, 50cm, 90cm and 125cm wavelengths
and achieved near simultaneous spectra. 
We fit synchrotron self absorption (SSA) and free-free models to the data. 
We compare the size of SN 
obtained using SSA fits to that of size extrapolated from VLBI measurements
at various epochs using public data at earlier epochs. 
We find  that the synchrotron self absorption process is insufficient to
reproduce the observed size of the supernova under the assumption of 
equipartition between magnetic fields and relativistic electrons. 
We also derive the evolution of spectral index and magnetic field 
at several epochs.

\end{abstract}

\section{Introduction}
\label{sec:1}
SN 1993J, a typical type IIb supernova,  exploded on March 28, 1993.
It was the nearest extragalactic supernova observed (3.63 Mpc away) 
and due to its
high positive declination, it was easily accessible by most of the 
telescopes for observations for most part of the year.
For this reason SN 1993J is the most detailed studied supernova 
after SN 1987A, in all
wavebands. 
The radio emission in supernovae is due to the synchrotron emission 
of relativistic electrons in the presence of magnetic field.
The radio emission is absorbed
in its early phases due to the presence of dense circumstellar medium.
It can also be absorbed because of its own highly dense ejecta (synchrotron
self absorption).

In section 2 we describe the GMRT observations. In section 3, we discuss
the spectra, the derived physical parameters from the various fits to the 
spectra  and their physical interpretations.                                                                                
\section{Observations}
                                                                                
We observed SN 1993J with Giant Meterwave Radio Telescope in 1420,
610, 325 and 235 MHz wave bands.
0834+555 and 1034+565 were used a phase calibrators.
3C48 and 3C147 were used as flux and 
bandpass calibrators.
The data was analyzed using the
Astronomy Image Processing Software (AIPS).
                                                                                
\begin{figure}
\begin{minipage}[t]{70mm}
\begin{center}
\includegraphics[height=7.5cm]{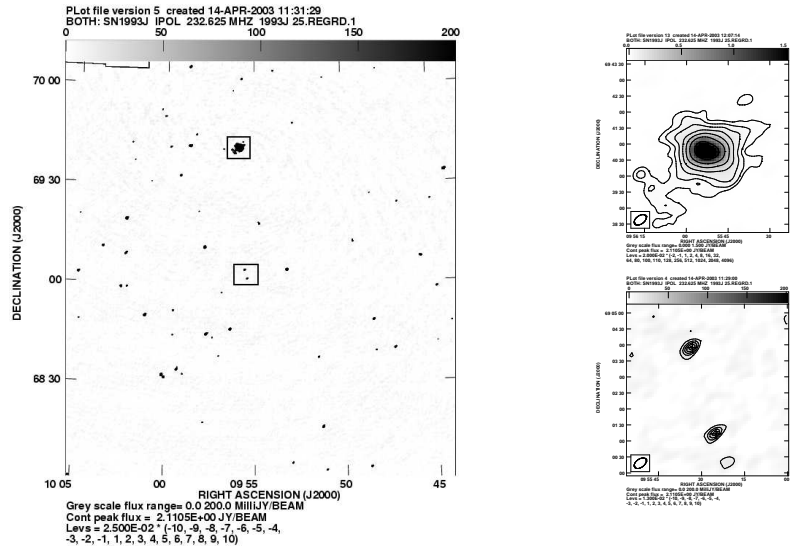}
\end{center}
\end{minipage}
\begin{minipage}[t]{50mm}
\begin{center}
\vspace*{-7.0cm}
\includegraphics[height=3.2cm]{poonamchandra_fig2.eps}
\includegraphics[height=3.2cm]{poonamchandra_fig3.eps}
\end{center}
\end{minipage}
\small\caption{ GMRT contour map of FOV 
of SN 1993J in 243 MHz band (Observed on Mar8,02). Upper panel is 
M82 and lower panel is M81 (Top left) and SN 1993J (Bottom right).
 \label{fig: 243_map}}
\end{figure}

Figure \ref{fig: 243_map} shows the FOV of SN 1993J at 243 MHz.
Right hand upper panel is the zoomed in plot of M82 (neighbouring galaxy in FOV)
 and lower panel is 
SN (bottom right) and its parent galaxy M81 (top left).  
                                                                                
\begin{figure}
\includegraphics[height=6.0cm,angle=270]{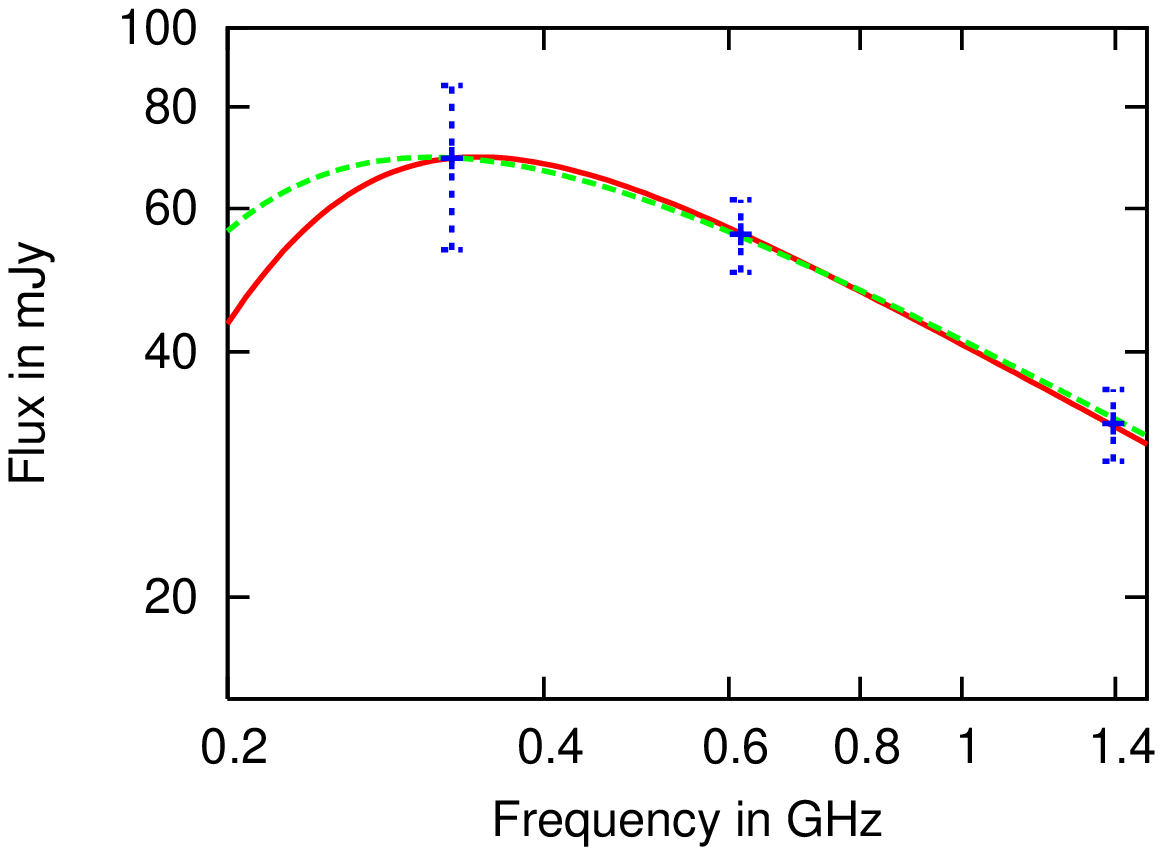}
\includegraphics[height=6.0cm,angle=270]{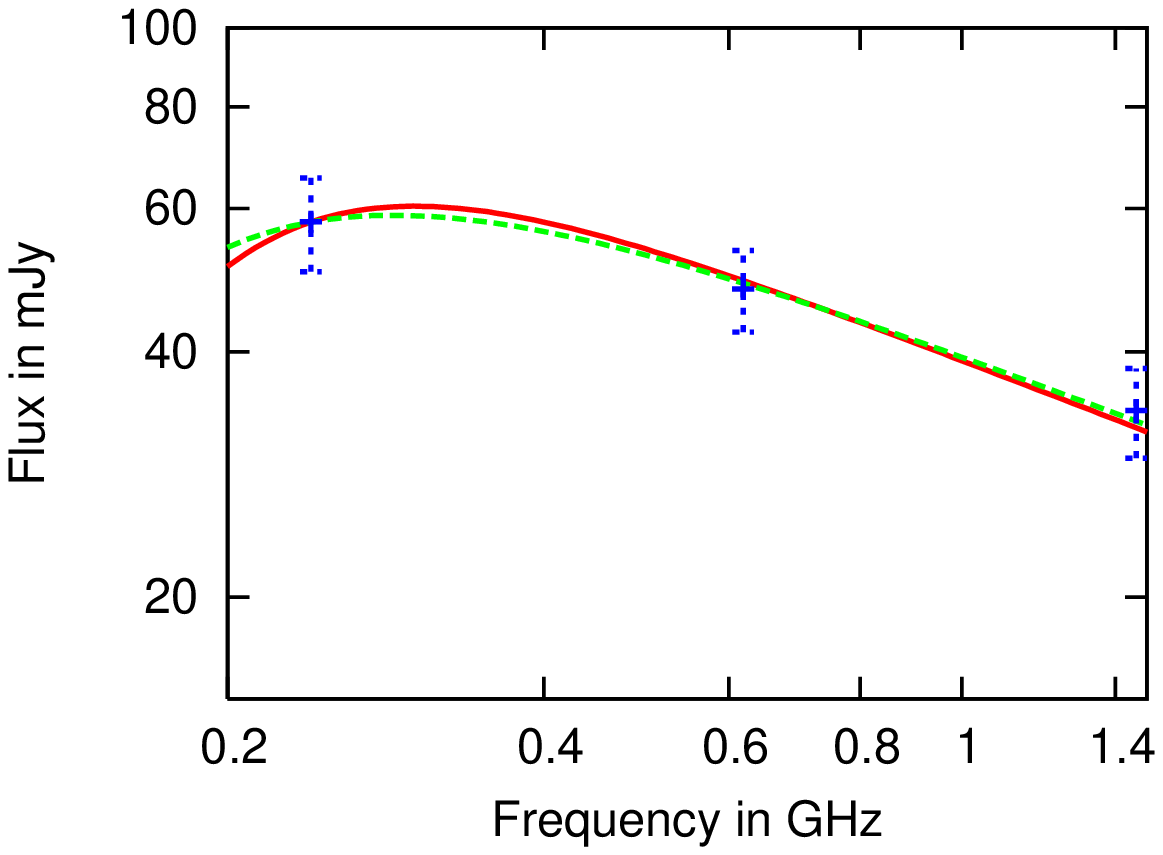}
\includegraphics[height=6.0cm,angle=270]{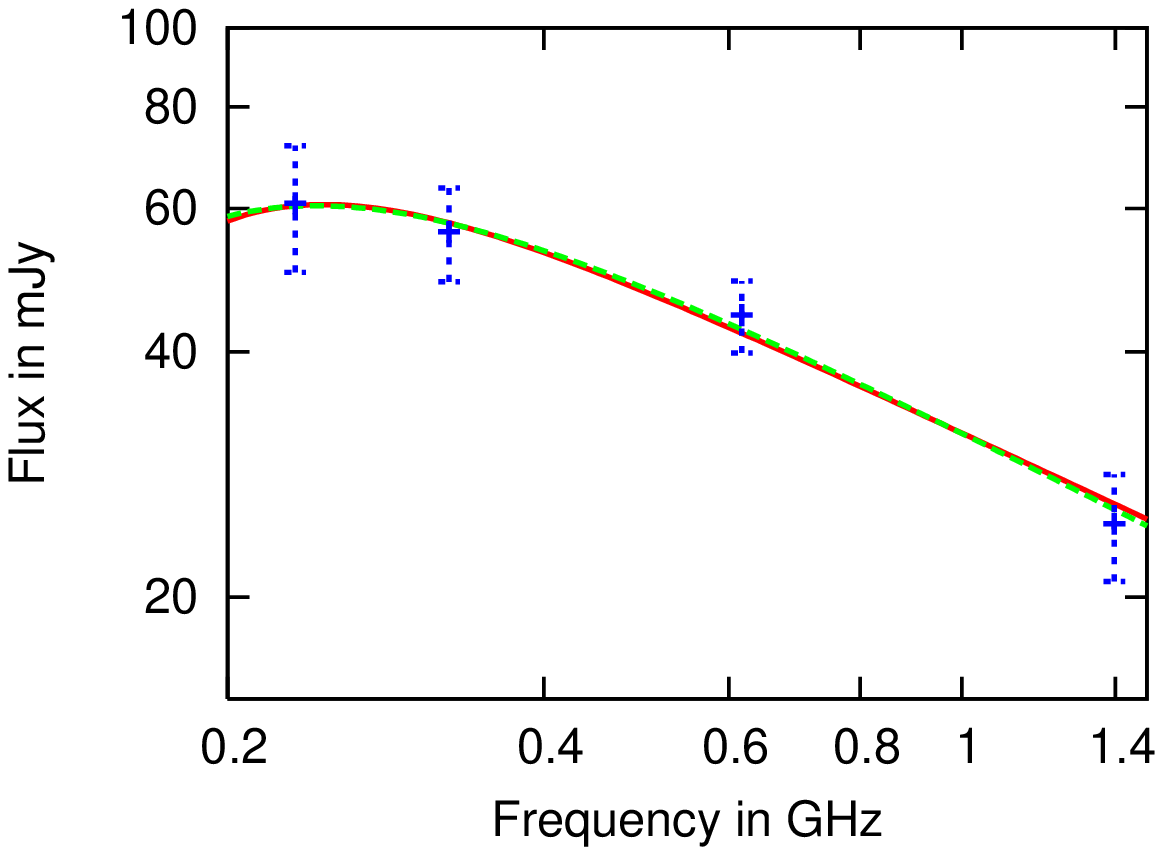}
\includegraphics[height=6.0cm,angle=270]{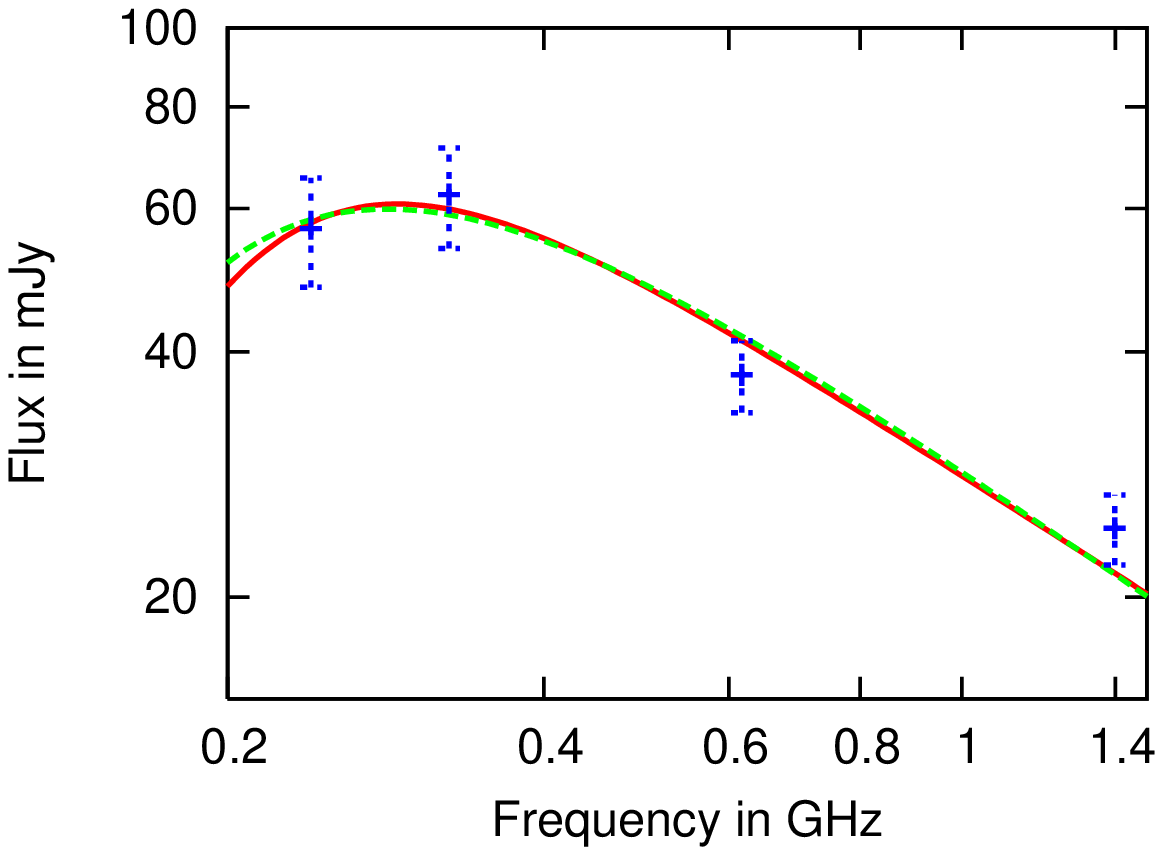}
\small\caption{Spectrum at four epochs (Clockwise $\sim$
8.2, $\sim$ 8.7, $\sim$ 9, and $\sim$ 9.5 years since explosion). 
Red (solid) line shows the SSA fit and green (dashed) shows the free-free fit.
\label{fig:spec}}
\end{figure}

The table \ref{tab: spec} below gives the details of the observations used 
in this paper.
We plot near simultaneous spectra at 4 epochs (See Figure \ref {fig:spec}).
Since the supernova is 10 years old, we do not expect its
flux to change by any significant amount in few days. 
The flux errors quoted in the table \ref{tab: spec} are
$$ \sqrt(( map~rms)^2 + (10\%~ of~ the~ peak~ flux~ of ~ SN)^2)$$
This 10\% of the peak flux error takes into account 
any possible systematic errors
in GMRT as well as any little variation in flux in the short time
difference due to near simultaneous spectra.
\begin{table}
\begin{center}
\caption{Details of observations of near simultaneous spectra of SN 1993J
\label{tab: spec}}
\begin{tabular}{lllllll}
\hline\noalign{\smallskip}
Date of  & Days since & Frequency & Time on & No. of  & Flux density & rms\\
Observation&  explosion & in GHz  &Source (Hrs) &   Antennae& mJy   & mJy \\
\noalign{\smallskip}\hline\noalign{\smallskip}
Jul 5,01        & 3022  & 0.327 & 2.5  & 18 & 69.2$\pm$15.8 & 2.5\\
Aug 24,01       & 3072  & 0.616 & 2    & 24   & 55.8$\pm$5.7  & 0.36\\
Jun 2,01        & 2988  & 1.393 & 3    & 28 & 32.67$\pm$3.3 & 0.19\\
\noalign{\smallskip}\hline\noalign{\smallskip}
Dec 31,01       & 3199  & 0.239 & 3    & 20 & 57.8$\pm$7.6 & 2.5\\
Dec 30,01       & 3198  & 0.619 & 1.7  & 20   & 47.8$\pm$5.5 & 1.9\\
Oct 15,01       & 3123  & 1.396 & 2    & 24   & 33.9$\pm$3.5 & 0.3\\
\noalign{\smallskip}\hline\noalign{\smallskip}
Mar 08,02& 3266 & 0.232   & 3.3   & 17 & 60.9$\pm$10.8 & 4.1\\
Mar 07,02& 3265 & 0.325   & 4.1   & 24 & 56.2$\pm$7.4  & 1.9 \\
Mar 08,02& 3266 & 0.617   & 3  & 25 & 44.4$\pm$4.5  & 0.32\\
Apr 07,02& 3296 & 1.396    &2.3   &25 & 24.6$\pm$3.7 & 1.0\\
\noalign{\smallskip}\hline\noalign{\smallskip}
Sep 16,02  & 3458& 0.240   & 3.4  &17  & 56.7$\pm$8.7 & 4.0\\
Aug 16,02& 3427 & 0.331   & 2  &16  & 62.4$\pm$8.8 & 2.7\\
Sep 16,02& 3458 & 0.617   & 2.7  & 15 & 37.5$\pm$3.8 & 0.4\\
Sep 21,02& 3463 & 1.397   & 2  &25  & 24.3$\pm$2.4 & 0.2\\
\noalign{\smallskip}\hline
\end{tabular}
\end{center}                                                                                
\end{table}

\begin{figure}
\centering
\includegraphics[height=10.3cm,angle=270]{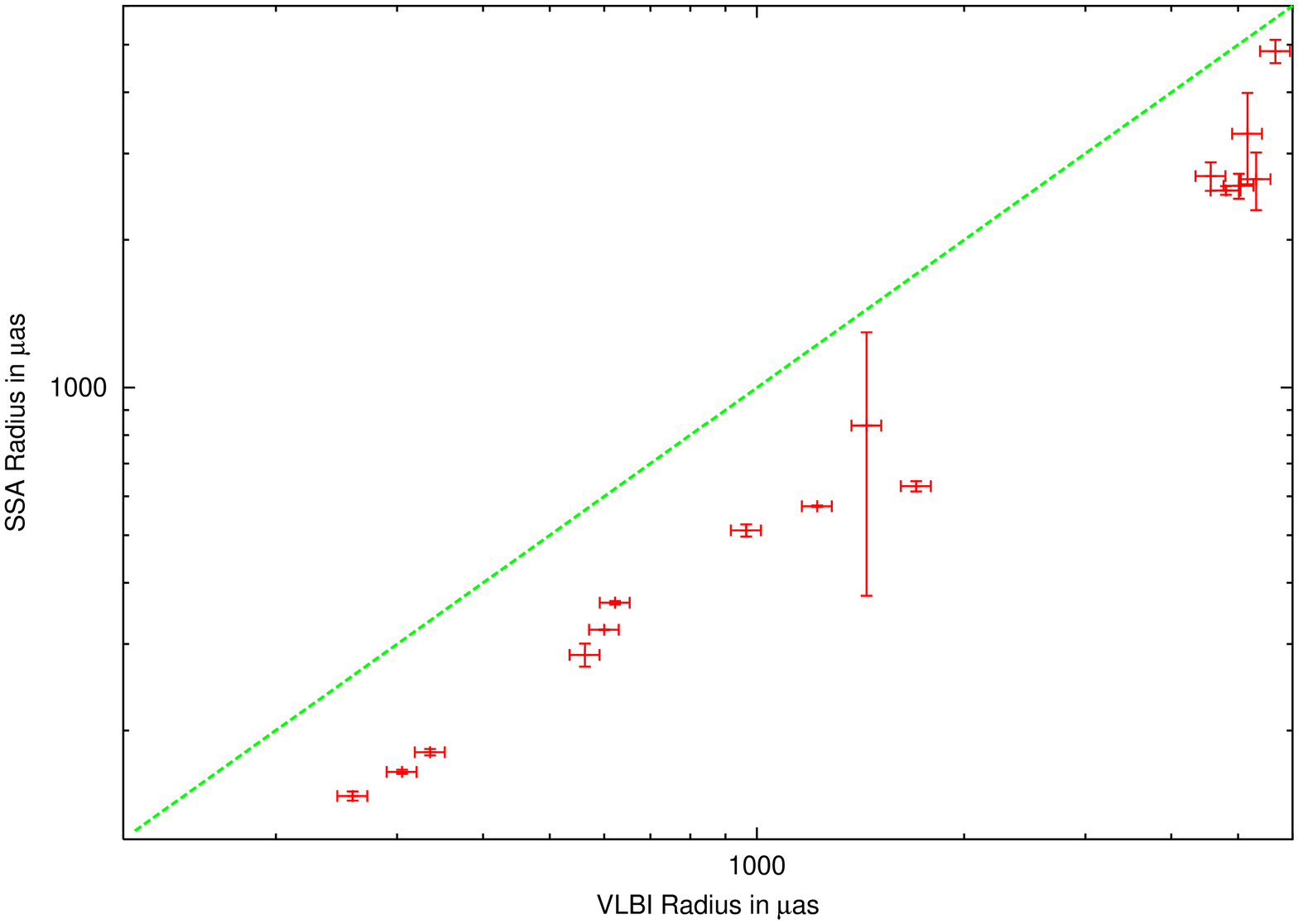}
\small\caption{The fig shows comparison between the VLBI radii of SNe
(\cite{bar02}) and the best fit synchrotron self absorption radius.
The straight line is the line where both the radii will be same. Note that
best fit SSA radius is much lower than VLBI radii. \label{fig: vlbi_ssa}}
\end{figure}
                                                                                
\section{Results and Interpretation}
                                                                                
We have fit the synchrotron self absorption models to the GMRT data and
got the best fit radius, magnetic field and spectral index. We
fitted the parameters under the assumption of equipartition
which seems to be a reasonable approximation \cite{sco77}.
Since in the two cases , we have only
3 datasets for the spectrum, we tried different values of spectral index and
get the best ${\chi}^2$ fit. We derived the size of the VLBI supernova
from \cite{bar02} and added 5\% errorbars to account for
any possible error in measurement. Figure \ref{fig: vlbi_ssa} shows
the plot of best fit synchrotron self absorption radius against
VLBI radius at various epochs. For early epochs,
we have used VLA public data and data from \cite{per02}. The green 
line is where
both the radii are same. It is noteworthy
that the synchrotron self absorption  radius is significantly smaller than the
VLBI radius at all epochs, contrary to findings of \cite{sly90}.
A combined SSA plus free-free fit also did not improve the results.

\begin{figure}
\includegraphics[height=5.9cm,angle=270]{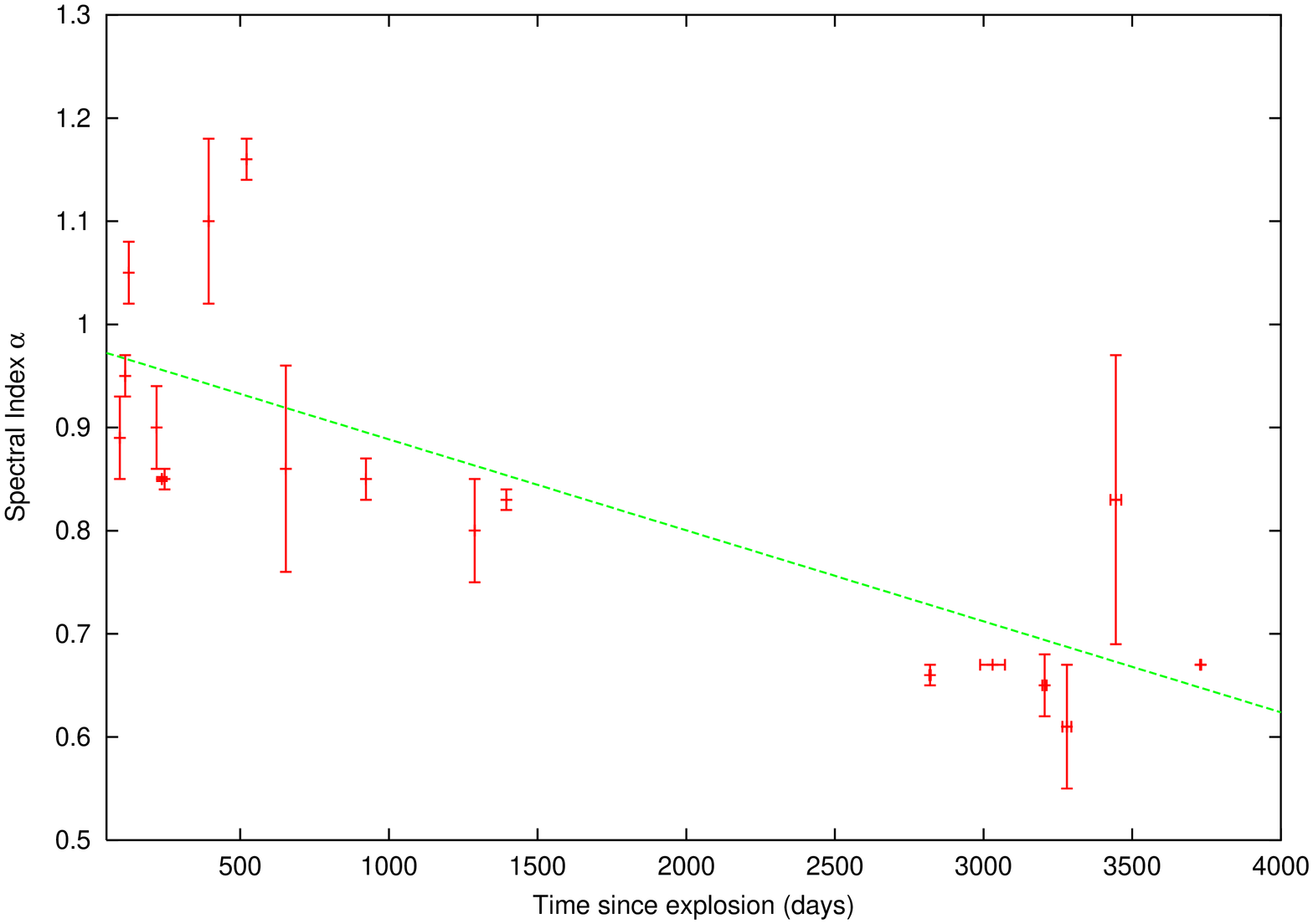}
\includegraphics[height=5.9cm,angle=270]{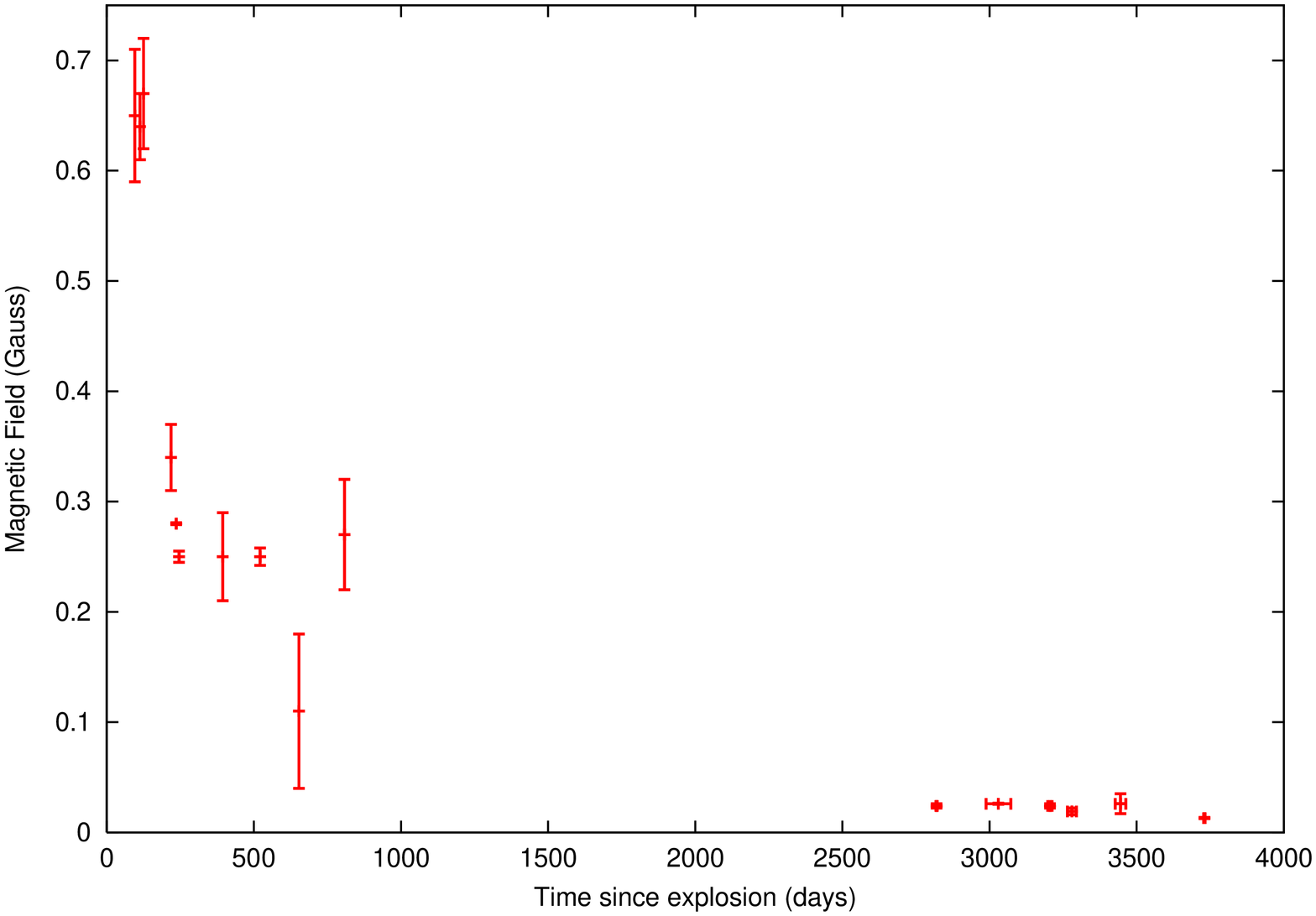}
\small\caption{The left panel shows the evolution of spectral index $\alpha$
with time. The right panel shows the evolution of the magnetic field.
\label{fig: alpha_B}}
\end{figure}

\begin{figure}
\includegraphics[height=4.9cm]{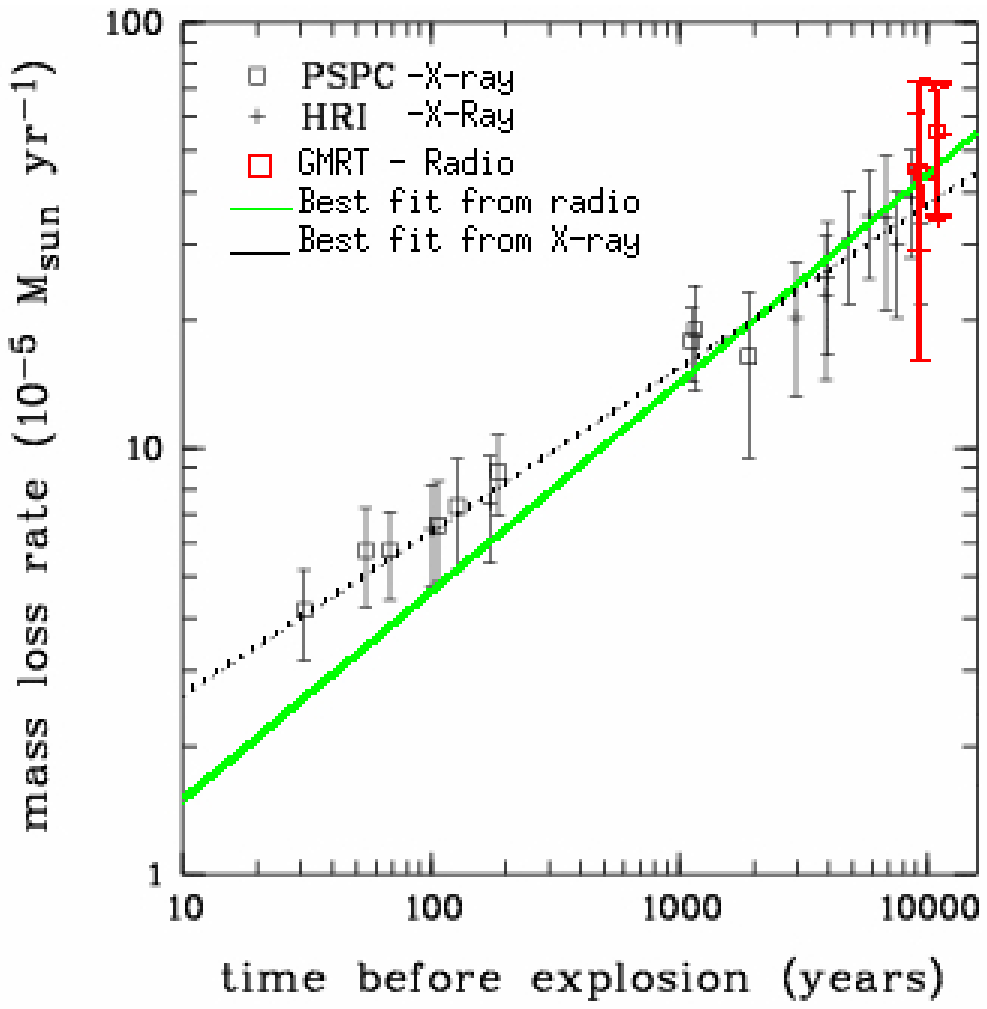}
\includegraphics[height=4.9cm]{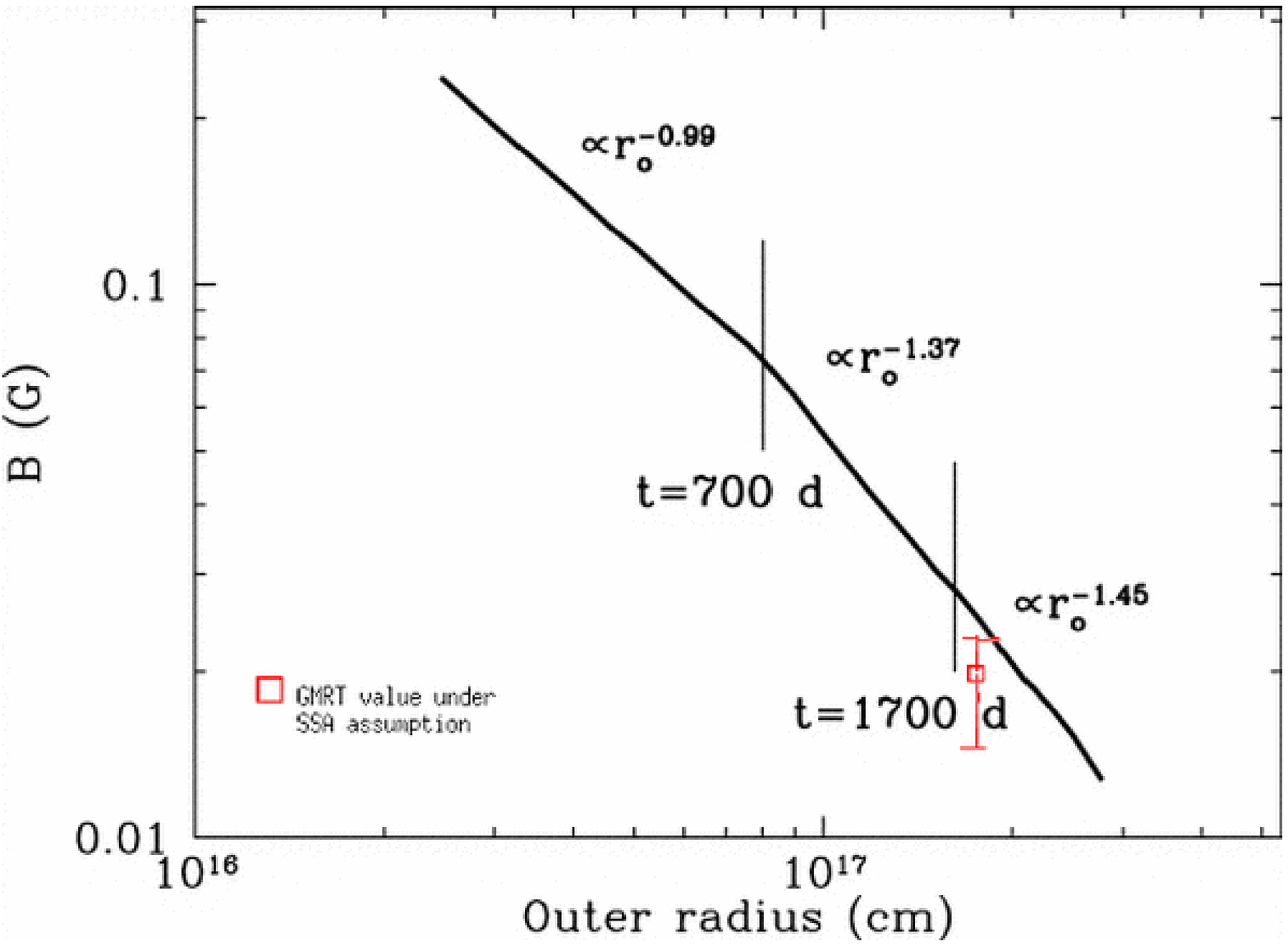}
\small\caption{Left panel is X-ray mass loss rates (black,\cite{imm01})
 and radio
mass loss rate (green,\cite{van94}). Red pts. are mass loss 
rates obtained from GMRT.
Right panel shows the mag. field derived from GMRT data (square) superimposed 
on top of Fig. 13 of \cite{bar02}.  
\label{fig: BandM}}
\end{figure}

Figure \ref{fig: alpha_B} shows the evolution of spectral index and magnetic
field with time. We find that the  spectrum of SN becomes flatter with time 
The magnetic field is also decreasing with
time, a reason for decreasing radio flux, because synchrotron emission
mechanism becomes less efficient as magnetic field decreases. 
We also calculate the mass loss rate for T$_e$=7$\times$10$^4$ K 
(assuming a constant temperature at these late epochs \cite{fra98}).
Figure \ref{fig: BandM} shows the radio mass loss rate 
curve (green,\cite{van94})
 overlaid on that of X-ray \cite{imm01} and  red points are mass loss 
rates obtained from GMRT 
data. Within error bars they seem to fit well with both radio and X-ray curves.
Our derived magnetic field given on the right panel of fig. \ref{fig: BandM}
is marginally consistent with that of Bartel \cite{bar02} under 
the assumpiton of
equipartition between magnetic field and relativistic particles.

We thank the staff of GMRT (NCRA-TIFR) that made these
observations possible.

%%%%%%%%%%%%%%%%%%%%%%%%%%%%%%%%%%%%%%%%%%%%%%%%%%%%%%%%%%%%%%%%%%%%%%

\printindex
\end{document}